\documentclass[%
reprint,
superscriptaddress,
amsmath,amssymb,
aps,
pra,
]{revtex4-2}

\usepackage{graphicx}
\usepackage{dcolumn}
\usepackage{bm}
\usepackage{bbm}

\usepackage{url}
\usepackage{xcolor}
\usepackage{amsthm}
\usepackage{slashed}
\usepackage{mathtools}
\usepackage{graphicx}
\usepackage{hyperref}
\usepackage{float}
\usepackage{comment}

\hypersetup{
  colorlinks   = true, 
  urlcolor     = blue, 
  linkcolor    = blue, 
  citecolor   = red 
}

\newcommand{\eqnref}[1]{Eq.~(\ref{#1})}
\newcommand{\figref}[1]{Fig.~\ref{#1}}
\newcommand{\sfigref}[2]{Fig.~\hyperref[#1]{\ref{#1}#2}}

\newcommand{\abs}[1]{\left| #1 \right|}

\newcommand{\beq}{\begin{equation}}
	\newcommand{\eeq}{\end{equation}}
\newcommand{\beqd}{\begin{equation*}}
	\newcommand{\eeqd}{\end{equation*}}
\newcommand{\dou}{\partial}
\newcommand{\up}{\uparrow}
\newcommand{\down}{\downarrow}
\newcommand{\expect}[1]{\left\langle #1 \right\rangle}
\newcommand{\bpm}{\begin{pmatrix}}
	\newcommand{\epm}{\end{pmatrix}}
\newcommand{\sx}{\sigma_{x}}
\newcommand{\sy}{\sigma_{y}}
\newcommand{\sz}{\sigma_{z}}
\newcommand{\tx}{\tau_{x}}
\newcommand{\ty}{\tau_{y}}
\newcommand{\tz}{\tau_{z}}
\newcommand{\id}[1]{\mathbbm{1}_{#1}}
\newcommand{\ot}{\otimes}
\newcommand{\td}{\tilde}
 
\newcommand{\mysub}[1]{\raisebox{-2pt}{$\scriptstyle #1 $}}

\DeclareSymbolFont{cmbrightop}{OT1}{cmbr}{m}{n}
\DeclareMathSymbol{\sfPsi}{\mathalpha}{cmbrightop}{9}

 \newcommand{\bcen}{\begin{center}}
 \newcommand{\ecen}{\end{center}}
 \newcommand{\btab}{\begin{tabular}}
 \newcommand{\etab}{\end{tabular}}
 \newcommand{\bdes}{\begin{description}}
 \newcommand{\edes}{\end{description}}

 \newcommand{\bea}{\begin{eqnarray}}
 \newcommand{\eea}{\end{eqnarray}}

 \newcommand{\half}{\frac{1}{2}}
 \newcommand{\bary}{\begin{array}}
 \newcommand{\eary}{\end{array}}
 \newcommand{\benum}{\begin{enumerate}}
 \newcommand{\eenum}{\end{enumerate}}
 \newcommand{\bitem}{\begin{itemize}}
 \newcommand{\eitem}{\end{itemize}}

 \newcommand{\Diel}{{K'}}
 \newcommand{\bsig}{{\boldsymbol{\sigma}}}

 \newcommand{\bphi}{\mbox{\boldmath $ \phi $}}
 \newcommand{\balp}{\mbox{\boldmath $ \alpha $}}
 
 \newcommand{\btau}{{\boldsymbol \tau}}

 \newcommand{\bLam}{\mbox{\boldmath $ \Lambda $}}
 
 \newcommand{\bDelta}{{\boldsymbol{\Delta}}}
 \newcommand{\bOne}{{\boldsymbol{1}}}
 \newcommand{\bk} { {\boldsymbol{k}} }

 \newcommand{\bmm} {{\boldsymbol{m}}}
 \newcommand{\bn} { \boldsymbol{n}}

 \newcommand{\br} { {\boldsymbol{r}}}

 \newcommand{\bv} { \mbox{\boldmath $v$}}

 \newcommand{\bJ} { {\boldsymbol{J}}}
 \newcommand{\bS} {\boldsymbol{S}}

 \newcommand{\mean}[1]{\langle #1 \rangle}

 \newcommand{\ket}[1]{{| #1 \rangle}}

 \newcommand{\calV}{\mathcal{V}}

 \newcommand{\Integers}{{\mathbb{Z}}}

 \newcommand{\ci}{\mathbbm{i}}

 \DeclareMathOperator{\tr}{tr}

\newcommand{\SMSymmetryMFT}{1}
\newcommand{\SMGrossNevue}{2}
\newcommand{\SMRG}{3}
\newcommand{\SMSchwinger}{4}

\newcommand{\SMSec}[1]{S#1}

\newcommand{\mytitle}{{Obstructed Atomic Insulators and Superfluids of Fermions Coupled to $\mathbb{Z}_2$ Gauge Fields}}

\begin{document}
	
	
	\title{\mytitle}
	
	\author{Bhandaru Phani Parasar}
	\email{bhandarup@iisc.ac.in}
	\affiliation{Centre for Condensed Matter Theory, Department of Physics, Indian Institute of Science, Bangalore 560012, India}
	\author{Vijay B.~Shenoy}
	\email{shenoy@iisc.ac.in}
	\affiliation{Centre for Condensed Matter Theory, Department of Physics, Indian Institute of Science, Bangalore 560012, India}
	
\begin{abstract}
We study spin-$\frac{1}{2}$ fermions coupled to $\mathbb{Z}_2$ gauge fields on a lattice. We show how a spatial modulation of the fermion hopping allows for the realization of various obstructed atomic insulators that host higher-order band topology. Studying the effect of quantum dynamics of the gauge fields within a simplified model, we find a rich phase diagram of this system with a number of superfluid phases arising from the attractive interactions meditated by the gauge fields. A key finding of this work is that the evolution from the Bardeen-Cooper-Schrieffer (BCS) superfluid state to a Bose-Einstein condensate (BEC) of tightly bound pairs occurs via the realization of these different superfluid phases separated by first-order transitions. 
\end{abstract}
\maketitle

\noindent
\underline{\em Introduction}: The discovery and classification of strong topological phases of non-interacting fermions\cite{KitaevPT,RyuTFW,HasanKane2010,QiZhang2011,ChiuRyu2016} marks an important milestone in condensed matter physics that has stimulated not only extensive theoretical investigations but also provide platforms for new technological realizations\cite{Collins2018}.
A conventional topological insulator in $d$ dimensions is insulating in the bulk and has gapless zero energy modes on the boundary of dimension $d-1$, often protected by symmetries \cite{KaneQSHE,KaneTI3D,KaneTIInv,MooreBalents2007,Roy2009}. Recent theoretical work has revealed that a more general characterization of topology can be realized in terms of obstructed atomic limits and how the interplay of crystalline symmetries brings about more possibilities\cite{Slager2013,BB1,BB2,ZZC,Josias}, leading to the notion of higher-order topological insulating (HOTI) phases\cite{Schindler}. An $n^{\text{th}}$ order topological insulator has symmetry-protected gapless  modes on the boundary of dimension $d-n$, with all higher dimensional boundaries being insulating with non-trivial topology. HOTIs have been realized experimentally in Bismuth \cite{Bismuth}, mechanical meta-materials \cite{meta}, acoustic systems \cite{acoustic1,acoustic2} and electric circuits \cite{circuit}. 

While the above developments are based on non-interacting band structures; an important concurrent direction is to investigate the consequences of these ideas in the presence of strong interactions and correlations. A fruitful way of studying strongly correlated systems\cite{Wegner1971,PESKIN1978,Dasgupta1981, Read1989,READ1989SUN,Fradkin1990,Hermele2004,Moessner2001,Baskaran1988,Affleck1988,Lee2006} such as quantum antiferromagnets, quantum dimer models, high $T_c$ superconductors etc.,~is to decompose the microscopic fermionic degrees of freedom into constituent partons and a gauge structure (since such decomposition is naturally endowed with some gauge redundancy). This leads to problems of partons coupled to dynamical gauge fields similar to those encountered in high energy physics and lattice gauge theories\cite{Kogut1979}, an understanding of which could throw light on many outstanding problems noted above.  Added to this are the possibilities  of realization of  such models of matter coupled to gauge fields in experiments \cite{Ohberg2011,Barbiero2019,Lewenstein2022}, which provides further impetus to this direction of study.

More recently, studying gauge fields coupled to charged fermionic matter has led to the identification of novel phases and phase transitions. The orthogonal metallic phase has been demonstrated in an exactly soluble model of fermions coupled to lattice gauge fields.\cite{NandkishoreOrthogonal} Using  Quantum Monte Carlo (QMC) formulation for $\mathbb{Z}_2$ lattice gauge coupled fermions free of the fermion sign problem,  ref.~\cite{Gazit1}  showed  that $\pi$ flux phase\cite{Lieb1994} with emergent Dirac fermions is spontaneously generated upon the increase of the fermion hopping amplitude. Tuning the quantum dynamics of the gauge fields, a continuous transition from the deconfined Dirac phase to confined BEC with the simultaneous onset of confinement of the gauge field and symmetry breaking is found.  Studies on closely related models \cite{Gazit2,Assaad,Maciejko,Tsvelik2020,PollmannMoroz2022} show several exotic phases and phase transitions. 

These developments motivate a broader question. Given that the Dirac semi-metallic phase attained at large hopping amplitudes is a gapless phase that typically occurs as a critical phase separating two phases of different patterns of entanglement, can the system be designed to promote interesting fermionic phases with different patterns of short-range entanglement? For example, can they produce obstructed atomic insulators? We address this question  by constructing and studying a model of fermions coupled to $\Integers_2$ gauge field that realizes obstructed atomic insulating phases.

The model we study in this paper has spatially varying hoppings of  fermions coupled to the gauge fields. Using a pattern of fermion hoppings as shown in \figref{fig:model}, we obtain a variety of phases, in the absence of quantum dynamics of the gauge fields, including  a metallic,  Dirac-semi metallic phase, trivial band insulators and obstructed atomic insulators when the strengths of the fermion hopping and the modulation of the hopping pattern are tuned. Several variants of obstructed atomic insulators, examples of higher-order topological insulators, are realized in the same model. We explore the effects of quantum dynamics of the gauge fields in a simplified model following reference \cite{Maciejko}, which captures the deconfined phase of the gauge fields. We show that turning on the dynamics of the gauge fields results in a rich phase diagram that includes a variety of superfluid/density-ordered phases, magnetic order, and valance bond solid phases with gapped (massive) fermions. We study the nature of the phase transitions between these phases using field theoretic techniques uncovering the difference between topologically trivial mass and     that which produces a HOTI phase. These results bring out the rich possibilities of studying a variety of topological phases within a single platform, which can potentially be realized in a cold atomic quantum simulator.

\begin{figure}
\includegraphics[width=0.66\columnwidth]{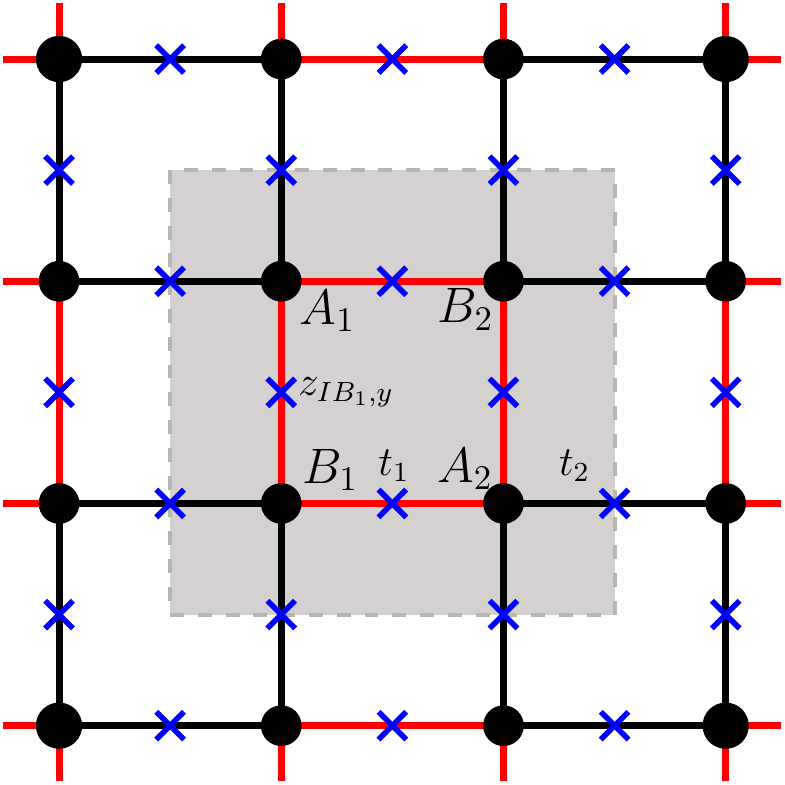}
\caption{Fermions coupled to gauge fields on a square lattice. The fermion hoppings are spatially varying with the pattern shown: red links are $t_1= t(1+r)$ and black links are $t_2 = t(1-r)$. The blue crosses show the gauge qubits. The shaded square shows the unit cell used in the analysis.
}
\label{fig:model}
\end{figure}

\noindent
\underline{\em Model}: We work on a square lattice with  a four-site basis consisting of sites $A_1,A_2,B_1,B_2$ as shown in \figref{fig:model}. At each site of the lattice, fermions are created by the operator $c^\dagger_{Ia\sigma}$ where $I$ is the unit cell index, $a \in \{A_1,A_2,B_1,B_2\}$, and $\sigma$ is a two-component spin (flavor) index. Along each link of this lattice, there is a gauge qubit whose Pauli-$Z$ operator is denoted by $Z_{Ia,\alpha}$, where $\alpha \in \{x,y\}$ indicates the direction of the link emanating from the site $Ia$. The fermions couple to the gauge fields via their hopping amplitudes $t_{Ia,\alpha}$ and the system is described by the Hamiltonian 
\begin{align}\label{eqn:model}
\nonumber
\mathcal{H'}=-&\sum_{Ia,\alpha, \sigma}\left(t_{\mysub{Ia, \alpha}}c_{\mysub{(Ia+\alpha)\sigma}}^{\dag}(Z_{\mysub{Ia\alpha}})^{q_\sigma}c_{\mysub{Ia\sigma}}^{}+\text{h.c.}\right) \\
&-\mu \sum_{I,a,\sigma} c^\dagger_{Ia\sigma} c_{Ia\sigma} -K\sum_{p}B_p -\Diel\sum_{I,a,\alpha}X_{\mysub{Ia,\alpha}}
\end{align}
where $(Ia + \alpha)$ is the site reached by traveling along the $\alpha$ link emanating from $Ia$, $t_{Ia,\alpha}$ is the hopping amplitude, $q_\sigma$ is a $\Integers_2$ valued charge (i.e., 0 or 1) of the fermion with  spin $\sigma \in \{\uparrow,\downarrow\}$, $\mu$ is the chemical potential, $p$ is a plaquette with $B_p = \prod_{(Ia,\alpha)/p}  Z_{Ia,\alpha}$ the plaquette magnetic term where the product is over the all links that touch the plaquette $p$,  $K$ is the inverse magnetic permeability of the gauge theory, $X_{Ia,\alpha}$ is the Pauli $X$ operator on the link $(Ia,\alpha)$, and $\Diel$ is the dielectric constant of the $\Integers_2$ gauge theory. The key aspect of this model is the spatially varying hopping amplitude where $t_{Ia,\alpha}$ takes on the value $t_1$ for links inside the unit cell shown in \figref{fig:model} and $t_{2}$ for links that cross from one unit cell to another. We parameterize $t_{1,2} = t(1\pm r)$ where $t$ is a scale of the gauge coupling (kinetic energy of the fermion), and $r$ is a dimensionless parameter. Throughout, we assume periodic boundary conditions in both $x$ and $y$ directions.

The model in \eqnref{eqn:model} has several important symmetries. There is a global U$(1)$ phase symmetry that corresponds to the conservation of the total number of particles.  If the $\Integers_2$ charges of the fermions are independent of their flavor, i.e., $q_\sigma = q$, then the system has a global SU$(2)$ symmetry  that acts on the spin labels. We assume this throughout this work.  Further, there is a local ``gauge symmetry'', in that, the unitary operators  \beq
G_{Ia} = A_{Ia} (-1)^{q n_{Ia}}
\eeq
transform the hamiltonian back to itself,
where $n_{Ia} = \sum_\sigma c^\dagger_{Ia\sigma} c_{Ia\sigma}$, and $A_{Ia} = \prod_{(Ia,\alpha)/Ia} X_{Ia,\alpha} $ where the product is over all the links that start or terminate at $Ia$. The physical Hilbert space of the theory is defined by that subspace where each $G_{Ia}$ acts as an identity, i.e., the Gauss law condition
\beq
G_{Ia} = 1, \;\;\; \forall Ia
\eeq
is imposed on all sites. Physically, this entails the absence of any external charges in the system other than those from the fermions and singularities of the gauge field itself.

If the fermions are not $\Integers_2$-charged (i.e., $q=0$), the ground state is a direct product of a fermionic Fermi-sea determined by the chemical potential with the ground state of the $\Integers_2$ gauge theory described by the last two terms in \eqnref{eqn:model}. The  $\Integers_2$ gauge theory is in a deconfined phase\cite{Tupitsyn2010} for $\Diel/K \lesssim 0.22$, and in a confined phase for larger values of $\Diel/K$. In the remainder of the paper, we focus on the more interesting case of $q=1$. We also restrict attention to the case where the filling of the fermions is one-half by tuning $\mu$ to a suitable value.

\noindent
\underline{$K'= 0$:} We first consider the case where the dielectric constant $\Diel$ of the gauge theory vanishes. Although this suppresses the quantum dynamics of the gauge fields, it reveals the physics that emerges from the competition between the kinetic energy of the fermions and the energetics of the plaquette magnetic fields.
 Noting that  $B_p, \forall p$ are conserved quantities when $\Diel=0$, we can write down the ground state wavefunction of the system for a particular set of values of $B_p$. Suppose the quantities $z_{Ia,\alpha} \in \pm 1$  (eigenvalue of the $Z_{Ia,\alpha}$ operator) describe the configuration of the gauge fields that realizes\footnote{In a system with periodic boundary conditions, there will be four distinct sets of $z_{Ia,\alpha}$ for a given set of $B_p$. These correspond to different values of the global Wilson line operators. On a finite system with periodic boundary conditions, one of the configurations will be chosen for the ground state in the presence of fermions. The one chosen will typically correspond with that configuring $z_{Ia,\alpha}$ whose Wilson line eigenvalues will be unity, and the fermions do not experience any flux enclosed along either direction of the torus.} the given set of $B_p$. The ground state is
\beq
\begin{split}
&\ket{\psi_{GS}\{z_{\mysub{Ia,\alpha}}\}} = \\
& \left(\prod_{Ia} \frac{\left(1+G_{\mysub{Ia}}\right) }{\sqrt{2}} \right) \left[ \left(\prod_{Ia\alpha}\ket{z_{\mysub{Ia,\alpha}}} \right)\otimes \ket{\text{FS}\{z_{\mysub{Ia,\alpha}}\}} \right]
\end{split}
\eeq
The ground state is attained by that configuration of $B_p$ that minimizes the energy. 

\begin{figure}
    \centering
    \includegraphics[width=\columnwidth]{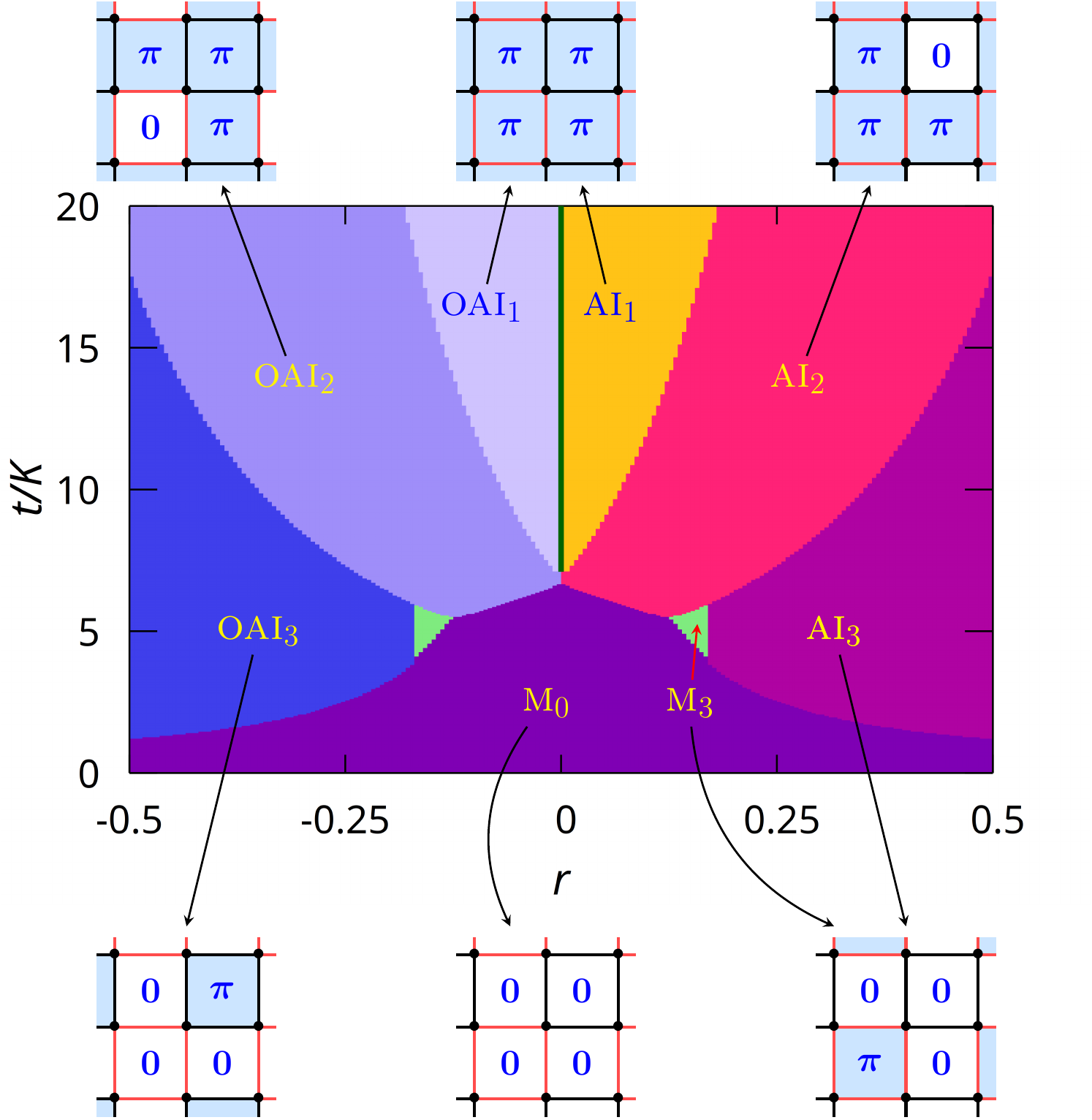}
    \caption{Ground state phase diagram of \eqnref{eqn:model} with $\Diel=0$. The phases labeled M are metals, AI are atomic insulators, OAI are obstructed atomic insulators. In each of these phases, the gauge field realizes the flux pattern as indicated by the dark arrows. }
    \label{fig:heq0pd}
\end{figure}

To determine the configuration of $B_p$ that minimizes the energy, we search among the configurations of $B_p$ that respect the translational symmetry associated with the unit cell shown in \figref{fig:model}. The ground states attained are shown in \figref{fig:heq0pd} as a function $t/K$ and $r$. The phase diagram consists of seven distinct phases. For small $|r| \lesssim 0.1$, the ground state is a metal with zero flux per plaquette. There is a first-order transition at a critical $t/K$ to a gapped phase called AI$_2$ for $r>0$ and OAI$_2$ for $r<0$. In the AI$_2$ phase obtained for $r>0$, all the plaquettes in the unit cell except the one that is enclosed by links with the black  $t_2$ hopping obtain a $\pi$-flux, while the OAI$_2$ gapped phase has $\pi$-flux in all plaquettes except that bounded by the red $t_1$ hoppings. At an even higher value of $t/K$, we see that gapped phases AI$_1 (r>0)$ and OAI$_1 (r<0)$ are obtained where a uniform $\pi$-flux is realized.  For larger values of $|r|$, $|r| \gtrsim 0.17$ an additional gapped phase (there is also an intervening metallic phase M$_3$ that appears), AI$_3$ ($r>0$) and OAI$_3$ ($r<0$), appears between the metal M$_0$ and the AI$_2$/OAI$_2$ phases.
For $r=0$, we find that the metal undergoes a transition to a gapped phase (AI$_2$ or OAI$_2$ depending on $r \to \pm 0$) at $t/K\approx6.72$ and remains in this phase until $t/K=6.922$, at which there is a first-order transition to a gapless Dirac semi-metallic phase with a uniform $\pi$-flux. This is consistent with ref.~\cite{Gazit1}, who, however, did not report the intervening gapped phase separating the metal and the Dirac semi-metal. We note that the Dirac semi-metal at $r=0$ has a diminished unit cell which encloses $\pi \left(\; mod \; 2\pi\right)$ $\Integers_2$ flux, whereas unitcell of OAI$_1$ or AI$_1$ ($r \neq 0$) has $0 \; \left(mod \; 2\pi\right)$ $\Integers_2$ flux. Thus the Dirac semi-metal and OAI$_1$ are distinct projective representations of the square lattice translation symmetry, and the band gap of OAI$_1$ must vanish as $r \to 0$. Whereas the diminishing of the unit cell does not occur for OAI$_2$ when $r \to 0$, indicating that it can remain gapped.


\begin{table}
\begin{centering}
\begin{tabular}{|c|c|c|c|c|c|}
\hline
 Phase & HSP &\#$\lambda_0$ & \#$\lambda_1$ &\#$\lambda_2$ & \#$\lambda_3$
\\
\hline
OAI$_{1/2/3}$ & \begin{tabular}{c}
     $\Gamma$ \\
     $M$\\
     $X$
\end{tabular}
& \begin{tabular}{c}
     $2$ \\
     $4$\\
     $8$
\end{tabular}& \begin{tabular}{c}
     $4$ \\
     $4$\\
     $8$
\end{tabular} &
\begin{tabular}{c}
     $6$ \\
     $4$\\
     $-$
\end{tabular} & \begin{tabular}{c}
     $4$ \\
     $4$\\
     $-$
\end{tabular} \\
\hline
AI$_{1/2/3}$ & \begin{tabular}{c}
     $\Gamma$ \\
     $M$\\
     $X$
\end{tabular} & \begin{tabular}{c}
     $4$ \\
     $4$\\
     $8$
\end{tabular}& \begin{tabular}{c}
     $4$ \\
     $4$\\
     $8$
\end{tabular} &
\begin{tabular}{c}
     $4$ \\
     $4$\\
     $-$
\end{tabular} & \begin{tabular}{c}
     $4$ \\
     $4$\\
     $-$
\end{tabular} \\
\hline
\end{tabular}
\caption{Number of irreducible representations of the $C_4$ rotations labelled by $\lambda_p=e^{i\frac{2\pi}{n}p}$ at different high symmetry points in the Brillouin zone. $n=4$ for $\Gamma$ and $M$, and $n=2$ for $X$. }
\label{table:symreps}
\end{centering}
\end{table}

The myriad phases obtained in this system raise some intriguing questions regarding their nature; for example, are the gapped phases similar or distinct? To understand the phases we first observe that all of the phases shown in \figref{fig:heq0pd} have a 4-fold rotational symmetry $C_4$, which may be seen by suitably enlarging the unit cell, which includes 16 sites. Next, we resolve the space of occupied fermionic states at high symmetry points $\Gamma = (0,0),M = (\pi,\pi)$ and $X=(\pi,0)$ in the Brillouin zone of the larger unit cell into (one-dimensional) irreducible representations of the $C_4$ rotations. The number of such states for each irrep for different points in the Brillouin zone is tabulated in table.~\ref{table:symreps}. In the AI phases obtained for $r>0$, the representations realized at the $\Gamma$ and $M$ points are the same -- such insulators have been termed as {\em atomic insulators}\cite{Benalcazar2019,Schindler2019} and hence the title AI for these phases.  On the other hand, remarkably, we see that the OAI insulators, all obtained when $r < 0$, have different representations at the different high-symmetry points and are realizations of {\em obstructed atomic insulators}\cite{Benalcazar2019,Schindler2019} and hence titled OAI. The obstructed atomic insulators are characterized by corner modes (whose energies lie in the band gap) when open boundary conditions are implemented and are characterized by a filling anomaly \cite{Benalcazar2019}. We have explicitly verified these points  in the systems on a finite-sized lattice with open boundary conditions.

A further natural question pertains to the distinction between different OAI phases. Interestingly, we find that all these are the same phase in that they carry the same classification data. In fact, we have found a way to connect the single particle hamiltonians of OAI$_2$ and OAI$_3$ through an adiabatic path (breaking time-reversal symmetry and not invoking the gauge fields)  that retains the gap throughout, demonstrating that the phase OAI$_2$ and OAI$_3$ are topologically indistinct. Note, however, that the flux patterns are distinct in these two phases, yet both of them realize the same fermionic band topology. 
It is remarkable that this system with but a few ingredients (fermions and $\Integers_2$ gauge fields) can produce such a rich phase diagram that could have many applications. For example, if the gauge theory described in \eqnref{eqn:model} arose from a partonic construction of a strongly correlated problem, then the physical fermions would be the product of the $c$-fermions and an Ising spin\cite{Sigrist2010,NandkishoreOrthogonal}. In this scenario, the interesting insulating phases such as OAI$_i$ $(i=1,2,3)$ are more appropriately called {\em orthogonal} obstructed atomic insulators, OAI$_i^*$, following reference \cite{NandkishoreOrthogonal}.

\noindent
\underline{$K'\ne 0$:} Next, we investigate the fate of these phases when the quantum dynamics of the gauge fields are turned on, i.~e., $\Diel\ne0$. In such a scenario,
the phase diagram  has to be evaluated with the recourse to quantum Monte-Carlo simulations as in \cite{Gazit1,Assaad}. Here we adopt a simpler approach following ref.~\cite{Maciejko}, which suggested coupling the fermions to the toric code \cite{Kitaev2004}. This entails replacing the dielectric term $-\Diel \sum_{Ia,\alpha} X_{Ia,\alpha}$ by $-h \sum_{Ia} A_{Ia}$. If this is achieved by a perturbation expansion of the $\Integers_2$ gauge theory (in the absence of the fermions), then $h \sim (\Diel)^4/K^3$, taking the gauge theory to the toric code limit, which is valid for $|\Diel/K| \ll 1$. This is the deconfined phase of the gauge theory. In this paper, we treat $h$ as an independent parameter that is allowed to take any real value. The key physical consequence of this formulation is that the gauge theory in the toric code limit is always in the deconfined phase. Although this approach cannot shed light on the confinement transition and its effect on the fermions, as shown below, it does display much interesting physics, even in the deconfined phase of the gauge theory.  

Within this simpler approach, we note that the Gauss law constraint in the singlet sector can be ``solved'' as
\beq
A_{\mysub{Ia}}=\left(-1\right)^{n_{\raisebox{-1pt}{$\scriptscriptstyle Ia$}}}=4\left(n_{\mysub{Ia,\up}}-\frac{1}{2}\right)\left(n_{\mysub{Ia,\down}}-\frac{1}{2}\right)
\eeq
leading us to the hamiltonian 
\beq
\begin{split}\label{eqn:modifiedmodel}
{\cal H} =-&\sum_{Ia,\alpha, \sigma}\left(t_{\mysub{Ia, \alpha}}c_{\mysub{(Ia+\alpha)\sigma}}^{\dag}(Z_{\mysub{Ia\alpha}})^{q_\sigma}c_{\mysub{Ia\sigma}}^{}+\text{h.c.}\right) \\
&-\mu \sum_{I,a,\sigma} c^\dagger_{Ia\sigma} c_{Ia\sigma}\\
&-K\sum_{p}B_p - 4 h \sum_{I a} \left(n_{\mysub{Ia,\up}}-\frac{1}{2}\right)\left(n_{\mysub{Ia,\down}}-\frac{1}{2}\right)
\end{split}
\eeq
which is the familiar Hubbard model\cite{Arovas2022}. Focusing on half-filling, we find that $\mu=0$. Further note that the system at this filling has an enlarged global symmetry SU$_{\textup{ph}}(2) \times$SU$_{\textup{sp}}(2) \sim $ SO$(4)$ which includes particle-hole  (ph) transformations and spin (sp) rotations\cite{Zhang1990}. In the analysis that follows, we use a version of the hamiltonian that makes this SU$_{\textup{ph}}(2) \times$SU$_{\textup{sp}}(2)$ symmetry manifest, see \cite{SM}(section \SMSec{\SMSymmetryMFT})  for details.

\begin{figure*}
    \centerline{
    \includegraphics[width=0.33\linewidth]{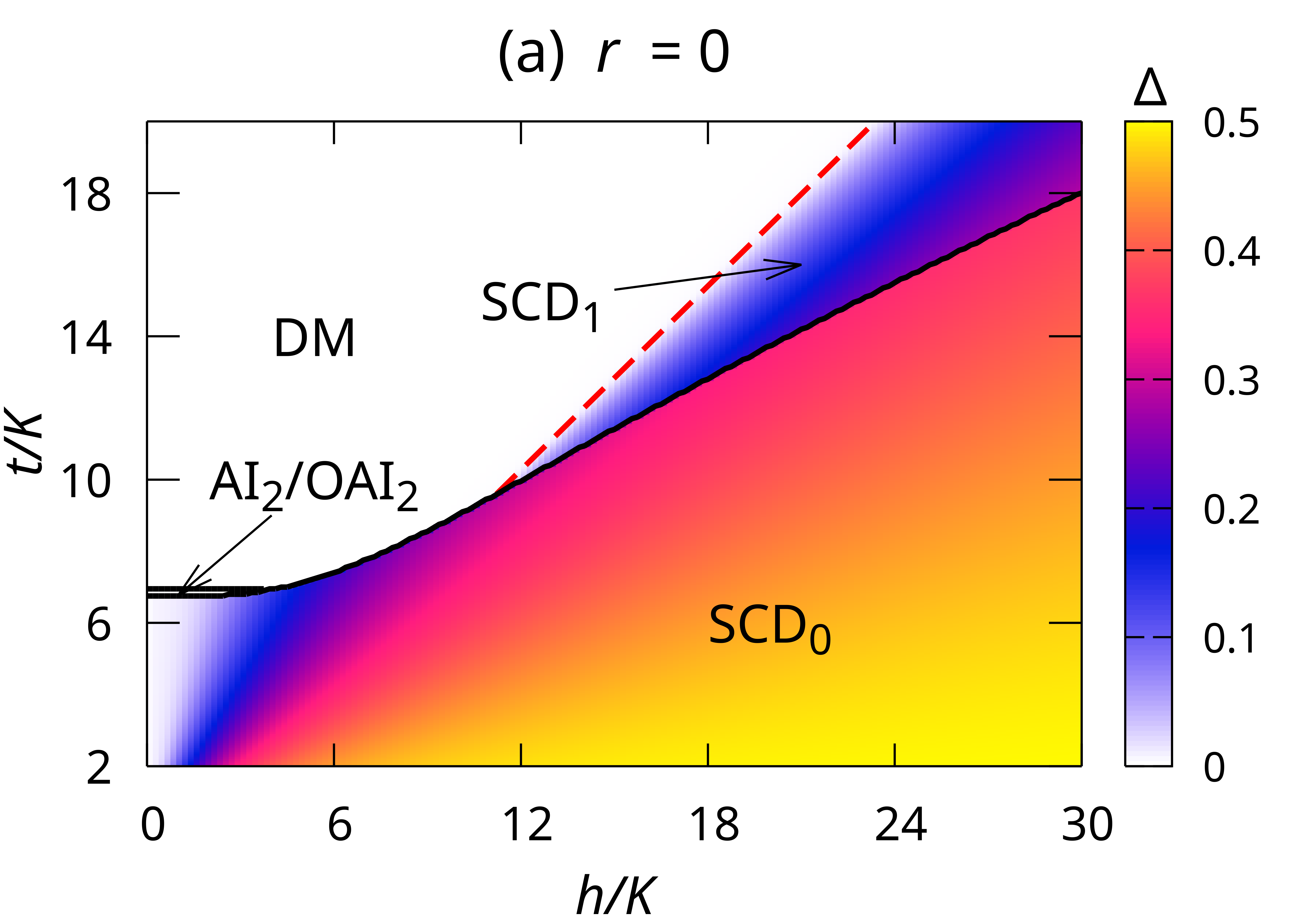}~
    \includegraphics[width=0.33\linewidth]{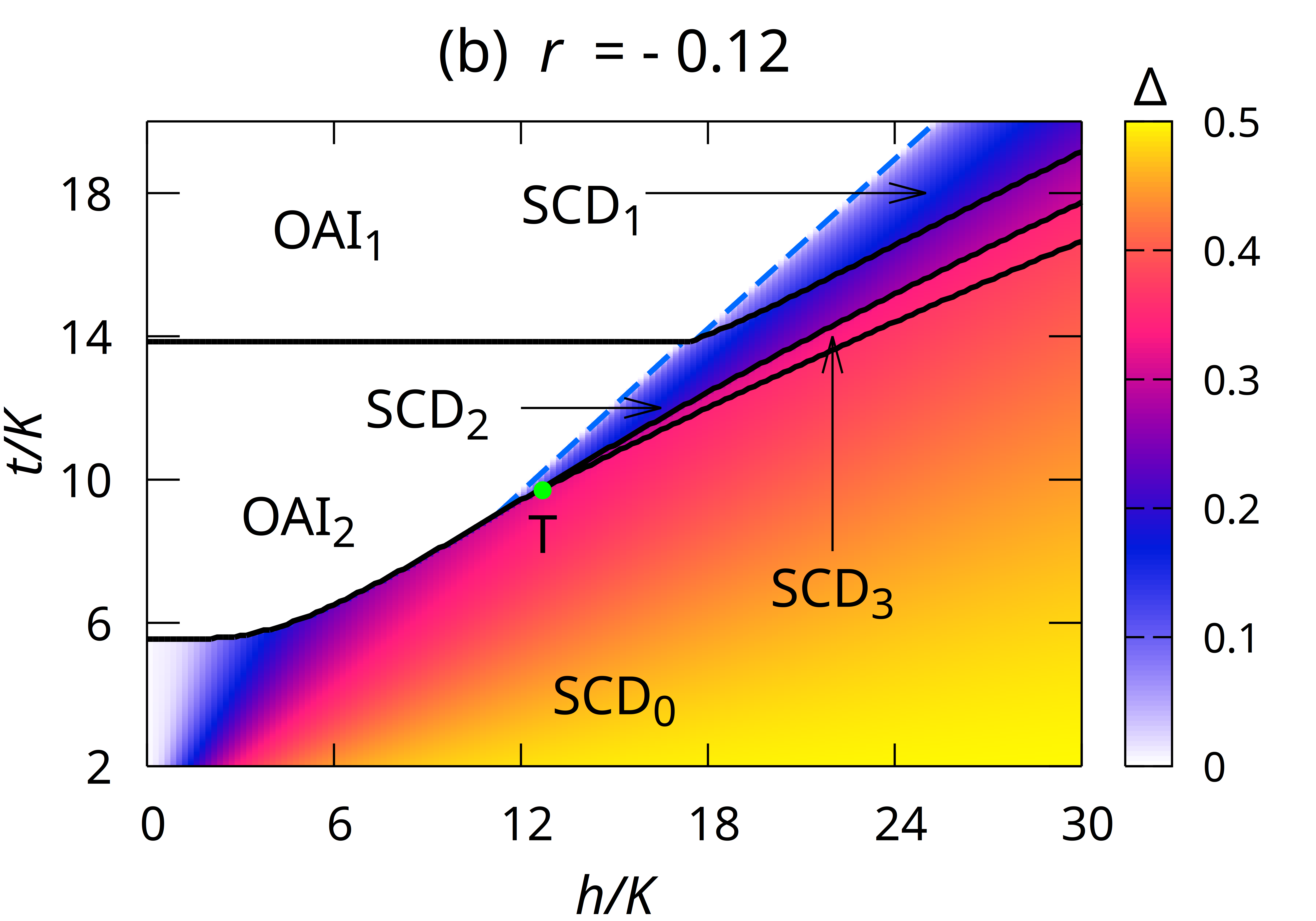}
    \includegraphics[width=0.33\linewidth]{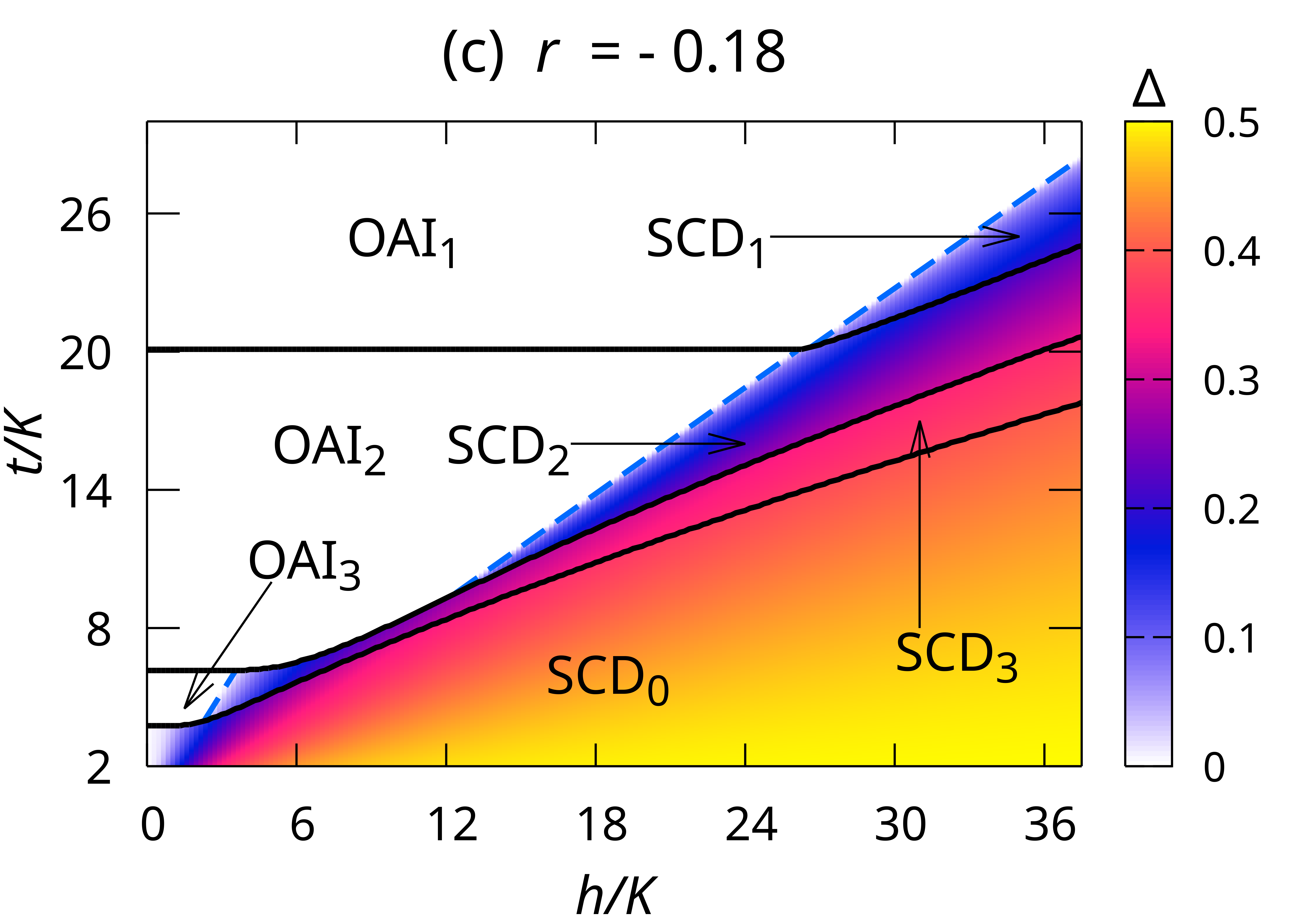}
    }
   
        \caption{Phase diagram of \protect{\eqnref{eqn:modifiedmodel}} obtained via meanfield theory for different values of $r$. The color shows the value of $\Delta$, the magnitude of the SCD order parameter. Solid black lines indicate first-order transitions. The dashed red line in panel (a) is a continuous transition line of the Gross-Neveu type. Dashed blue lines in panels (b) and (c) denote a continuous transition line of O(3)-$\phi^4$ universality. In (b), the bullet point in light green marked T denotes the triple point where SCD$_0$, SCD$_2$ and SCD$_3$ coexist.}
    \label{fig:mftpd}
\end{figure*}

\noindent
\underline{$h>0$:} We begin the discussion for the case where $h>0$ that corresponds to the attractive Hubbard model. 
The overall physics is well captured by a mean-field analysis which we adapt for the case here, taking care to note the SU$_{\textup{ph}}(2) \times$SU$_{\textup{sp}}(2)$ symmetry at half filling. For $h>0$, the SU$_{\textup{ph}}(2)$ symmetry is broken and the order parameter is described by a three-component vector (\cite{SM},section \SMSec{\SMSymmetryMFT}) with magnitude $\Delta$ which allows a transformation between the superconducting (SC) and $(\pi,\pi)$ density wave phase, which we collectively refer to as an ``SCD superfluid''. 
 When $h \ll t$, the metallic phase M$_0$ undergoes a BCS instability for any value of $h>0$ to an SCD phase with large pairs with $\Delta \ll 1/2$. For larger values of $h$, these pairs evolve into tightly bound bosons, and the superfluid  for $h \gg t$ can be viewed as a condensate of such bosons ($\Delta \approx \half$). When $r=0$, this is exactly the physics that we obtain for $t/K \leq 6.72$ as is shown in \figref{fig:mftpd}(a) where the superfluid phase obtained is labeled SCD$_0$ to denote the parent unpaired phase which has no flux per plaquette. For $6.72 \le t/K \le 6.92 $, the gapped OAI$_2$/AI$_2$ phase is stable for $h\ll t$, but undergoes a {\em first-order transition} to  the SCD$_0$ superfluid for larger values of $h$. For $6.72 < t/K < 10.0$, the Dirac metal (DM) obtained at $h=0$ is stable for finite $h$, and upon an increase of $h$, undergoes a first-order transition to the SCD$_0$ phase. For larger values of $t/K>10.0$, the DM phase itself undergoes {\em a continuous transition} (more on this below) to an SCD phase denoted by SCD$_1$ to indicate the uniform $\pi$-flux background of the parent normal-state. Interestingly, for larger values of $h$, the SCD$_1$ yields to the lower energy SCD$_0$ phase via a first-order transition.

 Matters put on an interesting hue when $r \ne 0$. For $r=-0.12$ as shown in \ref{fig:mftpd}(b), the insulator OAI$_2$ obtained at $t/K\gtrsim 5.5$  is stable to paring at small $h \ll t$, and upon the increase of $h$ undergoes a phase transition to the SCD$_0$ phase, via a first-order transition. However, for $10 \lesssim t/K \lesssim 14$, the OAI$_2$ undergoes a superfluid instability of its own and transits to an SCD$_2$ superfluid via a {\em continuous} transition\cite{Haldar2014}. Interestingly, the SCD$_2$ phase transforms to an SCD$_3$, a superfluid phase whose parent normal state is the OAI$_3$ phase! The SCD$_3$ phase, again, undergoes a first-order transition to an SCD$_0$ phase at larger values of $h$. An interesting aspect of the phase diagram is the presence of an SCD triple point ($t/K=9.7, h/K=12.7$)  where SCD$_{0}$, SCD$_2$ and SCD$_3$ phases coexist. For $t/K\gtrsim 14$, the OAI$_1$ phase stable at small $h$, undergoes a continuous phase transition to the SCD$_1$ phase (whose parent normal state is the OAI$_1$ phase). The phase eventually evolves to the SCD$_0$ phase, via {\em three first-order transitions}, first from SCD$_1$ to SCD$_2$, the second from SCD$_2$ to SCD$_3$, and the third from SCD$_3$ to SCD$_0$. There is an even richer phase diagram obtained for $r=-0.18$ where three distinct OAIs are realized at $h=0$. There are regimes of $t/K$ where each of these insulators OAI$_i$ ($i=1,2,3)$ undergoes a {\em continuous transition} to a SCD$_i$ phase; all of these evolve to the SCD$_0$ phase via a sequence of first-order transitions up on the increase of $h$. 
 
A key question that arises is the nature of the SCD$_i$ phases $(i=0,1,2,3)$. We have studied the dispersion of the Bogoliubov quasi-particles in each of these phases and concluded that all the SCD phases are topologically trivial. They also have an identical long-wavelength description in terms of an O(3) nonlinear-$\sigma$ model (without any topological term as the number of fermion flavors are even \cite{Abanov2000}) describing the long-wavelength fluctuations of the three component SDC order parameter whose amplitude modes are gapped. The difference between these phases will be found only in the cores of solitonic fields of the order parameter like skyrmions\cite{Abanov2000}, which may host localized fermionic modes. This provides an interesting line for future investigation.

We now investigate the nature of phase transitions between various phases. While many are first order, those between OAI$_i$ and SCD$_i$ ($i=1,2,3$) are continuous. The critical theory is described by an O(3) symmetric $\phi^4$ theory in $2+1$ dimensions \cite{Kleinert2001}. The most interesting continuous transition is the one between the DM phase to the SCD$_1$ phase. As detailed in \cite{SM}(section \SMSec{\SMGrossNevue}), this continuous transition can be modeled by a O$(N_\Sigma)$  Gross-Neveu theory \cite{GRACEY1990403,Gahemi,Boyack2021}, where $N_\Sigma =3$ in our case, described by gapless Dirac fermions and on O$(N_\Sigma)$ symmetric four-fermion term describing the interactions with a coupling constant $g$ (proportional to the parameter $h$). The physics of the masses that induce the OAI phase can be studied by including an additional mass term of the form $\bmm\cdot\bLam = \sum_b m_b \Lambda_b$ (where $\Lambda_b$ is a set of matrices that anti-commute with the gamma matrices). Performing a renormalization group analysis (\cite{SM}, section \SMSec{\SMRG}) to one loop order, we obtain the flow equations
\begin{align}\label{eqn:betafns}
\nonumber
s \dou_s g &= -\beta(g)=-\epsilon g+\frac{4N_{\Sigma}+N_{\gamma}-6}{\pi}g^2 \\
    s \dou_s m_b&=m_b\left(1+\frac{N_\Sigma}{\pi} g \right)
\end{align}
with $D=2+\epsilon$, where $D$ is the space-time dimension (3 in the present case), $N_\gamma = 8$, and $s \to \infty$ is the infrared limit. There are two fixed points. The first one that occurs at $(g=0,m_b=0)$ corresponds to the gapless Dirac theory, which is stable to small perturbations. The second one is the Gross-Nevue fixed point which is obtained at $(g = \frac{\pi \epsilon}{4 N_\Sigma + N_\gamma - 6}, m_b=0)$, obtains the critical interaction strength that destabilizes the Dirac fermions. We compute the anomalous dimensions of mass operators that produce OAI$_1$ phases and compare them with mass that produces a trivial gapped phase (setting $\Lambda_a$ as identity does this job) irrespective of the sign of the mass. We find that
\begin{align}\label{eqn:anomalousdim}
\nonumber
\eta_{\textup{OAI}_1}&=\frac{\epsilon N_\Sigma}{4N_{\Sigma}+N_{\gamma}-6}\\
\eta_{\textup{Trivial}}&=-\frac{\epsilon N_{\Sigma}}{N_\gamma-2}
\end{align}
($N_\Sigma=3, N_\gamma=8$ for the present case)
 It is interesting to note that these two mass terms have anomalous dimensions of opposite sign. 

We conclude the discussion of the $h>0$ phase diagram by noting that one obtains very similar physics for $r>0$ where one obtains a similar phase diagram involving SCD phases obtained by destabilizing AI$_{i}$ $(i=1,2,3)$ to obtain SCD'$_i$ phases.

\begin{figure}
\centerline{\includegraphics[width=\columnwidth]{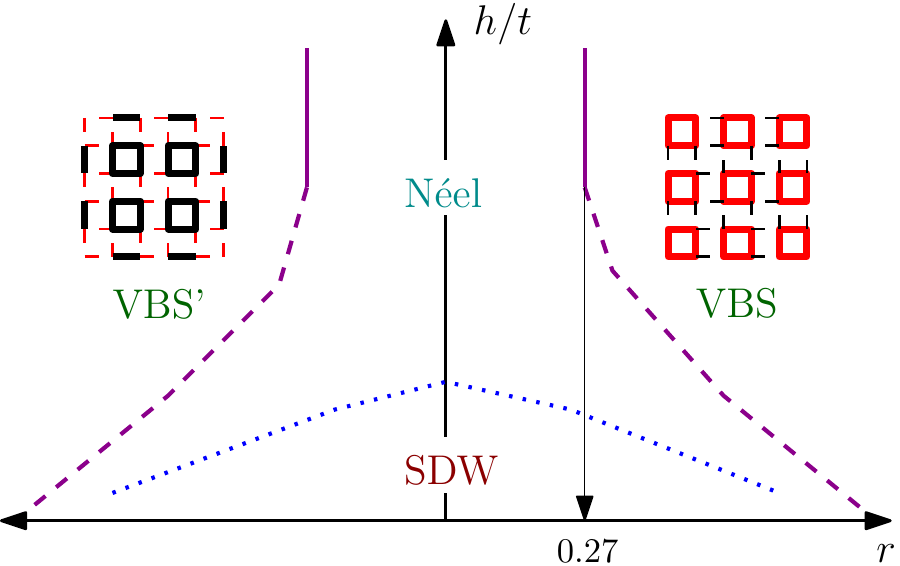}}
\caption{Schematic phase diagram for $h < 0$ (repulsive case). The phase boundary between the N\'eel state and the VBS states obtained at a large value of $h/t\gg1$ is obtained using a Schwinger boson mean-field theory where $r_c=0.27$. The dashed lines are schematic phase boundaries. The dotted line between the SDW (spin density wave state) and  N\'eel state represents a crossover. The region denoted by SDW can contain a rich structure with several phases like SDW$_i$ ($i=1,2,3)$ depending on the value of $t/K$. }
\label{fig:repulsive_scheme}
\end{figure}

\noindent
\underline{$h<0$:} The mean field phase diagram for $h<0$, which results in the repulsive Hubbard model, can be obtained by studying the symmetry breaking in the spin sector (preserving the particle-hole symmetry), which again leads to an O(3) vector order parameter (identifiable as the N\'eel order parameter). This again leads to spin-density wave SDW$_i$ (SDW'$_i$) phases obtained by destabilizing OAI$_i$(AI$_i$) phases ($i=1,2,3$). This mean-field analysis is, however, not reliable when $|h| \gg t$ where the system is a Mott insulator with forbidden double occupancy. The effective low-energy theory of the system becomes a Heisenberg model (irrespective of which insulator is the parent state) with ${\cal H}_{H} = -\sum_{Ia, I'a'} J_{Ia,I'a'} \bS_{Ia} \cdot \bS_{I'a'}$ where $J_{Ia,I'a'} \sim t_1^2/h$ on links inside the unit cell (\figref{fig:model}) and $J_{Ia,I'a'} \sim t_2^2/h$ on links across unit cells. For $r \sim 1$, the ground state for small $t/h$ is a valance bond solid (VBS) where the spins form resonating singlets on the links inside the unit cell. For $r \sim -1$, we obtain spins resonating on plaquettes bounded by dark-colored links in \figref{fig:model}. The state for $r=0$ is a N\'eel antiferromagnet (AF). We thus expect a transition from a N\'eel state to the valance bond state (cf. \cite{Takayama2001}) with the increase of $|r|$. The critical point $|r_c| = 0.27$  can be located using the Schwinger boson mean-field theory\cite{Auerbach,Sarker,auerbachbook} as detailed in \cite{SM}(section \SMSec{\SMSchwinger}). The full phase diagram (see \figref{fig:repulsive_scheme}) on the repulsive side involves several SDW$_i$ phases, which are smoothly connected to the phase, the details are left for future study.

\noindent
\underline{\em Concluding remarks}: This paper reveals the vast possibilities of realizing interesting phases in a system where fermions are coupled to gauge fields with spatially modulated hopping amplitudes. Our key results include the realization of various obstructed atomic insulators in these systems along with their instabilities. The study of the quantum dynamics of the gauge field leads to another interesting finding, i.e., the nature of the BCS to BEC cross-over in systems where fermions are coupled to gauge fields -- in such a system the crossover is much richer with many intervening phases. These findings will have not only interesting realizations in experiments in cold atomic systems but also stimulate further work in strongly correlated systems where parton decomposition techniques naturally lead to emergent gauge fields of the kind described here.

\noindent
The authors acknowledge support from SERB, DST, India via the CRG scheme.

\bibliographystyle{apsrev4-2}
\bibliography{ref}

\def\makeSM{1}
\ifdefined\makeSM

\newwrite\tempfile
\immediate\openout\tempfile=junkSM.\jobname
\immediate\write\tempfile{\thepage }
\immediate\closeout\tempfile

\clearpage
\newpage
\appendix

\renewcommand{\appendixname}{}
\renewcommand{\thesection}{{S\arabic{section}}}
\renewcommand{\theequation}{\thesection.\arabic{equation}}
 
\setcounter{page}{1}
\setcounter{figure}{0}

\begin{widetext}

\maketitle

\centerline{\bf Supplemental Material}
\centerline{\bf for}
\centerline{\bf \mytitle}
\medskip
\centerline{by Bhandaru Phani Parasar and Vijay B.~Shenoy}
\bigskip
\end{widetext}


\section{ SU$_{\textup{ph}}$(2)$\times$SU$_{\textup{sp}}$(2)  symmetry of the half-filled Hubbard model}\label{sec:PH_sym}

The Hubbard model at half filling on a generic bipartite lattice has an extra SU(2) symmetry\cite{Zhang1990,Arovas2022}. There are many ways to see this. First, consider
\beq\label{eqn:Hubbard}
\begin{split}
H = & \underbrace{-\sum_{I,J} \left(\td{t}_{IJ} c^\dagger_{IA\sigma} c_{J B \sigma} + \td{t}^*_{IJ} c^\dagger_{J B \sigma} c_{IA \sigma} \right)}_{H_K}\\
&- \underbrace{4h \sum_{I,a\in{A,B}} \left( n_{Ia\uparrow} - \half \right) \left(n_{Ia\downarrow} - \half \right)}_{H_I}
\end{split}
\eeq
The kinetic part of the hamiltonian includes hoppings only from sublattice $A$ to sublattice $B$ and vice-versa. Here, $\td{t}_{IJ}$ is the amplitude for hopping from $B$ sublattice of $J^{\text{th}}$ unitcell to $A$ sublattice of $I^{\text{th}}$ unitcell. The model has a global SU(2) in the spin space, generated by
\beq
S^i = \sum_{I a} c^\dagger_{I a \sigma} \tau^{i}_{\sigma \sigma'} c_{I a \sigma'}
\eeq
which satisfy the su(2) algebra. Now consider the objects,
\beq\label{eqn:Jph}
J^z_{Ia} = \half( n_{Ia} - 1), J^{-}_{I a} = (-1)^a c^\dagger_{I a \uparrow} c^\dagger_{I a \downarrow}, J^{+}_{I a} = (J^{-}_{I a})^\dagger 
\eeq
We now see the following by explicit calculation
\beq\label{eqn:phsu2}
[J^z_{Ia},J^{\pm}_{I a}] = J^{\pm}_{I a}, \;\; [J^+_{I a},J^{-}_{I a}] = 2 J^z_{I a}
\eeq
which is the su(2) algebra. We can construct ``global operators'' 
\beq\label{eqn:Global}
J^z= \sum_{I a }J^z_{I a}, \;\;\; J^{\pm}= \sum_{I a} J^{\pm}_{I a}
\eeq
(which again satisfies the su(2) algebra) to find
\beq
[J^z, H] = 0, [J^{\pm},H] = 0
\eeq
where in the last equation, we have used $\td{t}_{IJ} =\td{t}^*_{IJ}$ and  $c_{I \up} c_{I\down} (2 - n_{I \up} - n_{I \down}) = 0$ is the zero operator as it vanishes on 4 basis states at site $I$.
We thus see that in a system without a time reversal breaking magnetic field (where we can choose all hoppings to be real) we can get an additional SU$(2)$ symmetry that mixes particles and holes. 
To get an understanding of this symmetry, we need to look at the local Hilbert space which is made of four states, $\ket{0}, \ket{\uparrow}, \ket{\downarrow}, \ket{\uparrow\downarrow} \equiv \ket{D}$. While the spin SU$_{\textup{sp}}$(2) symmetry does  admixes the $\ket{\up}$ and $\ket{\down}$ states, the SU$_{\textup{ph}}$(2) admixes the $\ket{0}$ and $\ket{D}$ states.

To make both sp-SU(2) and SU$_{\textup{ph}}$(2) explicit, we first define
\beq
\psi^\dagger_{Ia} = \begin{pmatrix}
c^\dagger_{I a \uparrow} & c^\dagger_{I a \downarrow }
\end{pmatrix}
\eeq
The SU$_{\textup{sp}}$(2) is described by a unitary matrix
\beq
U(\theta,\bn) = e^{\ci \frac{\theta}{2} \bn \cdot \bsig} = \cos{\frac{\theta}{2}} \bOne + \ci \sin{\frac{\theta}{2}} \bn \cdot \bsig
\eeq
such that
\beq
\psi^\dagger_{I a} \to \psi^\dagger U(\theta,\bn) 
\eeq
To get the SU$_{\textup{ph}}$(2) in the mix, we observe that
\beq
\begin{pmatrix} c_{I a \uparrow} \\
c_{I a \downarrow}
\end{pmatrix} = 
\psi \to  U^\dagger(\theta,\bn) \psi
\eeq
under SU$_{\textup{sp}}$(2) transformation. Now, we want a ``annihilation operator like object'' that  transforms like $\psi^\dagger$. Consider $\psi^T$,
\beq
\psi^T  \to \psi^T U^*
\eeq
Now consider
\beq
\tilde{\psi} = \psi^T (-\ci \sigma_y)\;\; \tilde{\psi} \to \tilde{\psi} (\ci \sigma_y)U^* (-\ci \sigma_y) = \tilde{\psi} U
\eeq
We thus see that if we define an object
\beq
\sfPsi_{I a} := 
\begin{pmatrix}
c_{I a \downarrow} & -c_{I a \uparrow}\\
c^\dagger_{I a \uparrow} & c^\dagger_{I a \downarrow } 
\end{pmatrix}, \;\;\;\; 
\sfPsi^\dagger_{I a} = 
\begin{pmatrix}
c^\dagger_{I a \downarrow} & c_{I a \uparrow } \\
-c^\dagger_{I a \uparrow} & c_{I a \downarrow}
\end{pmatrix}
\eeq
which transforms under SU$_{\textup{sp}}$(2) as
\beq
\sfPsi_{Ia} \to \sfPsi_{I a} U(\theta,\bn), \;\; \sfPsi^\dagger_{Ia} \to U^\dagger(\theta,\bn) \sfPsi^\dagger_{I a}
\eeq
Next, we explore how the SU$_{\textup{ph}}$(2) defined by \eqnref{eqn:Jph} and \eqnref{eqn:phsu2} act on the fermion operators. To see this let us note
\beq
\begin{split}
[n_{I a},c^\dagger_{I a \sigma}] = \half c^\dagger_{I a \sigma}, [n_{I a},c_{I a \sigma}] = -\half c_{I a \sigma}
\end{split}
\eeq
and, defining $P^\dagger_{Ia} = c^\dagger_{I a \uparrow } c^\dagger_{I a \downarrow }$, $P_{Ia} = c_{I a \downarrow} c_{I a \uparrow}$
\beq
\begin{split}
[P^\dagger_{I a}, c^\dagger_{I a \sigma}] = 0, & \;\;\; [P_{I a}, c^\dagger_{I a \sigma}] = \sigma c_{Ia \bar{\sigma}} \\
[P^\dagger_{I a}, c_{I a \sigma}] = \bar{\sigma} c^\dagger_{I a \bar{\sigma}}  & \;\;\; [P_{I a}, c_{I a \sigma}] = 0
\end{split}
\eeq
which leads to
\beq
\begin{split}
    [J^z_{Ia}, c^\dagger_{I a \sigma}] = \half c^\dagger_{I a \sigma},  &\;\;\;\;  [J^z_{Ia}, c_{I a \sigma}] = -\half c_{I a \sigma}, \\
    [J^x_{I a},c^\dagger_{Ia \sigma}] = \frac{(-1)^a}{2} \sigma c_{I a \bar{\sigma}}, &\;\;\;\; [J^x_{I a},c_{Ia \sigma}] = \frac{(-1)^a}{2} \bar{\sigma} c^\dagger_{I a \bar{\sigma}}, \\
    [J^y_{I a},c^\dagger_{Ia \sigma}] = \frac{(-1)^a}{2 \ci} \bar{\sigma} c_{I a \bar{\sigma}}, &\;\;\;\; [J^y_{I a},c_{Ia \sigma}] = \frac{(-1)^a}{2 \ci} \bar{\sigma} c^\dagger_{I a \bar{\sigma}}. \\
\end{split}
\eeq
Now consider the unitary operator using \eqnref{eqn:Global}
\beq\label{eqn:Vdef}
\calV(\phi,\bmm) = e^{\ci \phi \bmm \cdot \bJ}
\eeq
We now explore how the fermion operator transforms under an infinitesimal $\calV$ ($\phi$ is small).
\begin{widetext}
Clearly,
\beq
\begin{split}
&e^{\ci \phi \bmm \cdot \bJ_{Ia}} c^\dagger_{I a \sigma} e^{-\ci \phi \bmm \cdot \bJ_{Ia}} \approx  c^\dagger_{I a \sigma} + \ci \phi [\bmm \cdot \bJ, c^\dagger_{I a \sigma}] \\
 &= c^\dagger_{I a \sigma}  + \frac{\ci \phi}{2} \left[ (-1)^a m_x (\sigma c_{I a \bar{\sigma}})  + (-1)^a (\ci m_y) (\sigma c_{I a \bar{\sigma}})  + m_z c^\dagger_{I a \sigma} \right] \\
 &= c^\dagger_{I a \sigma}  + \frac{\ci \phi}{2} \left[ (-1)^a (m_x +\ci m_y) (\sigma c_{I a \bar{\sigma}})   + m_z c^\dagger_{I a \sigma} \right] 
\end{split}
\eeq
and similarly,
\beq
\begin{split}
&e^{\ci \phi \bmm \cdot \bJ_{Ia}} (\sigma c^\dagger_{I a \bar{\sigma}}) e^{-\ci \phi \bmm \cdot \bJ_{Ia}} \approx \\
&= \sigma c_{I a \bar{\sigma}} - \frac{\ci \phi}{2} \left[ 
(-1)^a (m_x - \ci m_y) (\sigma \bar{\sigma} c^\dagger_{I a \sigma}) + m_z \sigma c_{Ia\bar{\sigma}} \right] \\
&= \sigma c_{I a \bar{\sigma}} + \frac{\ci \phi}{2} \left[ 
(-1)^a (m_x - \ci m_y) c^\dagger_{I a \sigma} - m_z \sigma c_{Ia\bar{\sigma}}
\right] 
\end{split}
\eeq

We thus see that
\beq
\begin{pmatrix}
\sigma c_{I a \bar{\sigma}}\\
c^\dagger_{I a \sigma} 
\end{pmatrix}
\to
\left[ 
\begin{pmatrix}
1 & 0 \\
0 & 1
\end{pmatrix} +
\frac{\ci \phi}{2}
\begin{pmatrix}
m_z & (-1)^{a} \left(m_x - \ci m_y \right) \\
 (-1)^{a} \left(m_x + \ci m_y \right) & -m_z
\end{pmatrix}
\right]\begin{pmatrix}
\sigma c_{I a \bar{\sigma}} \\
c^\dagger_{I a \sigma} 
\end{pmatrix}
\eeq
\end{widetext}
Finally, we obtain
\beq
\begin{split}
\calV(\phi,\bmm)\sfPsi_{I a} \calV^\dagger(\phi,\bmm) &= V_a(\phi,\bmm) \sfPsi_{Ia} \\
V_a(\phi,\bmm) = e^{\ci \frac{\phi}{2} \bmm_{a} \cdot \btau}
\end{split}
\eeq
where $\btau$ are the Pauli matrices in ph-space and
\beq
\bmm_a = ((-1)^a m_x, (-1)^a m_y, m_z).
\eeq
Clearly,
\beq
\sfPsi^\dagger_{I a} \to \sfPsi^\dagger_{I a} V^\dagger_{a}(\phi,\bmm).
\eeq
Now the appearance of $(-1)^a$ in the definition of $\bmm_a$ owes to the definitions in \eqnref{eqn:Jph}. To understand why such a definition of the symmetry is warranted, we check the invariance of the hamiltonian under the symmetry operation \eqnref{eqn:Vdef}. We start with casting the hamiltonian \eqnref{eqn:Hubbard} in a more natural fashion
\beq
H_K = \half \sum_{IJ} -\td{t}_{IJ} \tr{\left(\sfPsi^\dagger_{IA} \tau_z \sfPsi_{JB} + \sfPsi^\dagger_{JB} \tau_z \sfPsi_{IA} \right)}
\eeq
First of all, SU$_{\textup{sp}}$(2) symmetry is immediate due to the cyclic invariance of the trace. Next, we note that
\beq
\begin{split}
& \calV(\phi,\bmm)\sfPsi_{I a} \calV^\dagger(\phi,\bmm) = \\
\half \sum_{IJ} & -t_{IJ} \tr{\left(\sfPsi^\dagger_{IA} V^\dagger_A(\phi,\bmm) \tau_z V_B(\phi,\bmm)\sfPsi_{JB} \right. } \\ &\left. + \sfPsi^\dagger_{JB}V^\dagger_{B}(\phi,\bmm) \tau_z V_A(\phi,\bmm)\sfPsi_{IA} \right)
\end{split}
\eeq
It can now be explicitly verified that
\beq
\tau_z V_B(\phi,\bmm) = V_A(\phi,\bmm) \tau_z
\eeq
leading to the invariance of $H_K$ under the action of $\calV$. We thus realize that  for real hoppings $t_{IJ}$, the hopping operator in ph-space looks like $\tau_z$. This is the origin of the  definition of $J$s in \eqnref{eqn:Jph}.

To cast this in an invariant form, we define
\beq
\Phi_{I a} = \frac{\sfPsi_{Ia} \sfPsi^\dagger_{I a} - \bOne}{2} =  \begin{pmatrix}
\half(1 - n_{Ia}) & P_{Ia} \\
P^\dagger_{Ia} & -\half(1 - n_{Ia}) 
\end{pmatrix}
\eeq
and obtain 
\beq
H_I = -4h\sum_{Ia} \left( \frac{1}{3} \tr \Phi_{I a}^2 - \frac{1}{4} \right)
\eeq
Spin SU(2) invariance follows from the invariance of $\Phi_{I a}$ under SU$_{\textup{sp}}$(2) transformation, and  SU$_{\textup{ph}}$(2) invariance follows from the cyclic invariance of the trace.

We thus see that the symmetry SU$_{\textup{ph}}$(2)$\times$SU$_{\textup{sp}}$(2) is manifest where the hamiltonian is written as
\beq
\begin{split}
H &= \half \sum_{IJ} -\td{t}_{IJ} \tr{\left(\sfPsi^\dagger_{IA} \tau_z \sfPsi_{JB} + \sfPsi^\dagger_{JB} \tau_z \sfPsi_{IA} \right)} \\
& - 4 h \sum_{Ia} \left( \frac{1}{3} \tr \Phi_{I a}^2 - \frac{1}{4} \right)
\end{split}
\eeq
Quite usefully, the form of the hamiltonian is also amenable to mean-field analysis. Consider $h>0$, then we can write
\beq\label{eqn:mftdecouple}
\Phi_{Ia}^2 =  \bDelta_{I a} \Phi_{I a} + \Phi_{I a} \bDelta_{I a}  - \bDelta_{I a} \bDelta_{I a} 
\eeq
where 
\beq\label{eqn:mftansatz}
\bDelta_{I a} = \mean{\Phi_{Ia}} = 
\begin{pmatrix}
\Delta_{Ia}^{(3)} & \Delta^{1}_{Ia} - \ci \Delta^{(2)}_{Ia} \\
\Delta^{(1)}_{Ia} + \ci \Delta^{(2)}_{Ia} & - \Delta_{Ia}^{(3)} 
\end{pmatrix}
\eeq
is a trace free Hermitian matrix with real $\Delta^{(i)}_{Ia}$. We now adopt the following anazatz
\beq
\Delta^{(1)}_{I a} = \Delta^{(1)}, \Delta^{(2)}_{I a} = \Delta^{(2)}, \Delta^{(3)}_{I a} = (-1)^{a} \Delta^{(3)} 
\eeq
where $(\Delta^{(1)} + \ci \Delta^{(2)} )$ is the superconducting order parameter, while $\Delta^{(3)}$ is the charge density wave order parameter. 

Suppose we have a meanfield ground state, then we can obtain another degenerate ground state by means of the symmetry operation $\calV(\bphi,\bmm)$. Under this operation
\beq
\bDelta_{Ia} \to V^\dagger_{a}(\phi,\bmm) \bDelta_{Ia} V_{a}(\phi,\bmm)
\eeq
which implies that the 3-vector
\beq
\begin{pmatrix}
\Delta^{(1)} \\
\Delta^{(2)} \\
\Delta^{(3)} 
\end{pmatrix} \to R(\phi,\bmm) \begin{pmatrix}
\Delta^{(1)} \\
\Delta^{(2)} \\
\Delta^{(3)} 
\end{pmatrix}
\eeq
where $R(\phi,\bmm)$ is the O(3) rotation matrix that corresponds to $\phi$. We thus see that we can ``rotate'' between superconductivity and charge density wave in a continuous fashion. 

For our problem, what this means is that we are breaking the SU$_{\textup{ph}}$(2) symmetry, and not merely the U$(1)$ of particle number. 

In our model, we have two sites each per sub-lattices $A$ and $B$. When $h>0$, mean field decoupling is performed in SC $+$ CDW channel (\eqnref{eqn:mftdecouple}). Defining the Nambu fermion object $\Psi_I$ according to 
\beq
\nonumber
\Psi_{I}^{T}=
\bpm c_{\mysub{IA_1 \up}}\!\!\!\!& c_{\mysub{IA_2 \up}}\!\!\!\!& c_{\mysub{IB_1 \up}}\!\!\!\!& c_{\mysub{IB_2 \up}}\!\!\!\!& c_{\mysub{IA_1 \down}}^{\dag}\!\!\!\!& c_{\mysub{IA_2 \down}}^{\dag}\!\!\!\!& c_{\mysub{IB_1 \down}}^{\dag}\!\!\!\!& c_{\mysub{IB_2 \down}}^{\dag} \epm
\eeq
We can write the Hubbard term in the mean-field approximation as
\begin{align}\label{eqn:MfHamScd}
\nonumber
H_{\text{I,MF}}=-\frac{8h}{3}&\sum_{I}\Psi_{I}^{\dag} \left(\Delta^{(1)} \xi_1 + \Delta^{(2)} \xi_2 + \Delta^{(3)} \xi_3\right) \Psi_I\\
&+\frac{8h}{3}\sum_{Ia} \left(\bDelta \cdot \bDelta\right)
\end{align}
where 
\beq\label{eqn:scdmasses}
\begin{split}
\xi_1&=\tx \ot \id{2} \ot \id{2}\\ 
\xi_2&=\ty \ot \id{2} \ot \id{2}\\ 
\xi_3&=-\tz \ot \sz \ot \id{2}
\end{split}
\eeq
$\xi_1$,$\xi_2$ are s-wave superconducting masses and $\xi_3$ is the mass corresponding to charge density wave.

Let us find the action of SU$_{\textup{ph}}$(2) symmetry on the object $\Psi_I$. We know that under the ph-symmetry transformation:
\beq
\nonumber
\bpm -c_{\mysub{Ia \up}} \\ c_{\mysub{Ia\down}}^{\dag} \epm \to V_a \left(\phi,\bmm \right) \bpm -c_{\mysub{Ia \up}} \\ c_{\mysub{Ia\down}}^{\dag} \epm
\eeq
Which can also be written as
\beq
\nonumber
\bpm c_{\mysub{Ia \up}} \\ c_{\mysub{Ia\down}}^{\dag} \epm \to \tz V_a \left(\phi,\bmm \right) \tz  \bpm c_{\mysub{Ia \up}} \\ c_{\mysub{Ia\down}}^{\dag} \epm
\eeq

Then it follows that the ph-symmetry transformation acts on the Nambu object $\Psi_I$ as
\beq
\Psi_I \to e^{i \frac{\phi}{2} \bmm \cdot \bv} \Psi_I
\eeq
where the generators are
\beq
\label{eqn:phsymgen}
\begin{split}
    v_1&=-\tx \ot \sz \ot \id{2} \\
    v_2&=-\ty \ot \sz \ot \id{2} \\
    v_3&=\tz \ot \id{2} \ot \id{2}
\end{split}
\eeq
A similar analysis can be performed for $h<0$ decoupling the four-fermion term in the N\'eel antiferromagnetic channel, which again provides for an O(3) order parameter.
\section{O$(N_\Sigma)$ Gross-Neveu model arising as the low energy theory of symmetry breaking in Dirac Metal}\label{sec:GNmodel}
In this section, we will be concerned about the effective theory for the electrons when the gauge fields are in $\pi$ flux state since we want to describe the critical properties of the continuous phase transition from the $\pi$ flux Dirac metal to the SCD$_1$ phase. The $\pi$ flux gauge fields give rise to emergent Dirac fermions. The Dirac kinetic hamiltonian below is written in the gauge shown in \figref{fig:piflux}:
\begin{figure}
    \centering
    \includegraphics[width=0.66\columnwidth]{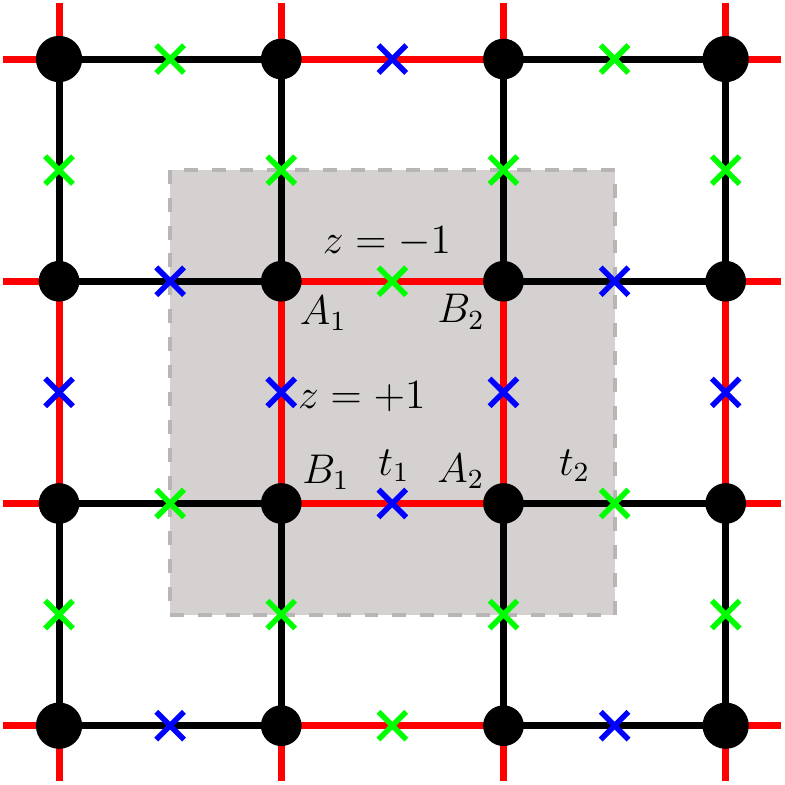}
    \caption{Configuration of $\Integers_2$ gauge fields for uniform $\pi$ flux through each plaquette. Links attached with blue and green crosses have $z=+1$ and $z=-1$, respectively.}
    \label{fig:piflux}
\end{figure}
\beq\label{eqn:piblochHam}
H_{\text{K}}=\sum_{\mysub{\bk\sigma}}c_{\mysub{\bk a\sigma}}^\dagger h^{ab}(\bk)c_{\mysub{\bk b \sigma}}
\eeq
where
\beq
\resizebox{0.9\linewidth}{!}{$ h(\bk)=
\bpm 0 & 0& t_1-t_2 e^{ik_y} & -t_1+t_2 e^{-ik_x} \\ 0 & 0 & t_1-t_2e^{ik_x}& t_1-t_2e^{-ik_y} \\ t_1-t_2 e^{-ik_y} & t_1-t_2e^{-ik_x} & 0 & 0 \\ -t_1+t_2 e^{ik_x} & t_1-t_2e^{ik_y} & 0&0 \epm$
}
\eeq
with $t_1=t\left(1+r\right)$, $t_2=t\left(1-r\right)$. The low energy hamiltonian is
\beq\label{eqn:DiracHam}
H_{\text{K}}=\sum_{\sigma} \int_{\abs{\bk}<\Lambda} d^2 \bk\,\, \psi_{\sigma}^{\dag}(\bk)\left(at \balp' \cdot \bk +2 \sqrt{2}rt \beta'\right)\psi_{\sigma}(\bk)
\eeq
where (The mass $\beta'$ gives rise to OAI$_1$ for $r<0$.)
\beq
\nonumber
\begin{split}
    \alpha_x'&= \sy \ot \sx \\
    \alpha_y'&=\sy \ot \sz\\
    \beta'&=\frac{1}{\sqrt{2}}\left(\sx \ot \id{2}+\sy \ot \sy\right)
\end{split}
\eeq
In terms of the Nambu spinor $\Psi(\bk)$,
\beq\label{eqn:DiracNambu}
H_{\text{K}}=\int d^2  \bk\,\, \Psi^{\dag}(\bk) \left(at \balp \cdot \bk +2\sqrt{2}rt\beta \right)\Psi(\bk)
\eeq
\beq
\begin{split}
\alpha_x&=\tz \ot \sy \ot \sx\\
\alpha_y&=\tz \ot \sy \ot \sz\\
\beta&=\frac{1}{\sqrt{2}}\left(\tz \ot \sx \ot \id{2}+\tz \ot  \sy \ot \sy\right)
\end{split}
\eeq
The SCD masses in \eqnref{eqn:scdmasses} and the $\alpha$ matrices satisfy
\begin{align}\label{eqn:algebra}
\nonumber
\{\xi_i,\xi_j \}&=2\delta_{ij}\;\;\;i,j=1,2,3\\
\{\alpha_{x,y},\xi_i\}&=0
\end{align}
Also, $\{\beta,\xi_i\}=\{\beta,\alpha_{x,y}\}=0$.
Now, we note that symmetry generators of the particle-hole transformation \eqnref{eqn:phsymgen} are commutators of the SCD masses. i.e.,
\beq
v_i=-\frac{\ci}{4}\epsilon_{ijk}[ \xi_j, \xi_k]
\eeq
Thus, the fermion bilinears $\Psi^{\dag}(\br)\xi_i \Psi(\br)$ transform as components of a $3$-vector under the particle-hole symmetry transformations. We can construct a scalar by taking the sum of squares of these bilinears. This gives a four-fermion term that respects the O$(3)$ particle-hole symmetry of the Hubbard term. Hence, it is reasonable to replace the Hubbard term with this Gross-Neveu term to study the critical properties of the phase transition.
\beq\label{eqn:GNscd}
H=H_{\text{K}}-h'\int d^2 \br\sum_{i}\left(\Psi^{\dag} \xi_i \Psi\right)^2
\eeq
We have the same situation for N\'eel AF order in a repulsive Hubbard model. Let us consider a generalization of the hamiltonian in \eqnref{eqn:GNscd}:
\begin{align}\label{eqn:gen_ham}
\nonumber
H=\int d^2 \bk\;\; \Psi^{\dag}\left(\alpha_x k_x+\alpha_y k_y + \sum_{m}m_b \beta_b\right)&\Psi\\
-u\int\;d^2 \br \sum_{i}\left(\Psi^{\dag}\xi_i \Psi\right)^2&
\end{align}
Here, $\beta_b$ is the set of masses in the single-particle hamiltonian and $\xi_i$ is the set of mass matrices entering the four-fermion interaction. There is no mass common to these sets. They all anticommute with each other and also gap out the Dirac fermions. i.e., they satisfy the algebra
\begin{align}\label{eqn:gen_algebra}
\nonumber
\{\alpha_{x,y},\beta_b \}&=0\\
\nonumber
\{\beta_b,\beta_{b'}\}&=2\delta_{b b'}\\
\nonumber
\{\alpha_{x,y},\xi_i \}&=0\\
\nonumber
\{\xi_i,\xi_j\}&=2\delta_{ij}\\
\{\xi_i,\beta_b\}&=0
\end{align}
The Euclidean action, upon taking $\gamma_0=i \alpha_x \alpha_y$ and making the change of variables $\Psi^* \rightarrow \overline{\Psi} \gamma_0$, $\Psi \rightarrow  \Psi$ is
\begin{align}\label{eqn:gen_action}
S=\int\;\; d^3x \Bigg[\overline{\Psi} \left(\gamma^\mu \dou_\mu +\sum_{b}m_b \Lambda_b\right)\Psi \Bigg. 
\Bigg. -u \sum_{i}\left(\overline{\Psi}\Sigma_i \Psi\right)^2\Bigg]
\end{align}
with
\beq
\begin{split}
    \gamma_0&=i \alpha_x \alpha_y\\
    \gamma_1&=-i \gamma_0 \alpha_x=-\alpha_y\\
    \gamma_2&=-i \gamma_0 \alpha_y=\alpha_x\\
    \Sigma_i / \Lambda_b&=\gamma_0 \xi_i / \gamma_o \beta_b
\end{split}
\eeq
We have
\beq
\begin{split}
   \{\gamma_\mu,\gamma_\nu\}&=2\delta_{\mu\nu}\\
   \{\Sigma_i,\Sigma_j\}&=2\delta_{ij} \\
   \{\Lambda_b,\Lambda_{b'}\}&=2\delta_{b b'} \\
   [\gamma_\mu,\Sigma_i / \Lambda_m]&=0
\end{split}
\eeq
Let us denote the number of masses entering the interaction term as $N_\Sigma$ and the dimension of $\gamma^\mu$ matrices as $N_\gamma$. The internal SO$(N_\Sigma)$ symmetry transformations are generated by
\beq\label{eqn:gen_symgen}
Y_{i j}=\frac{\ci}{2}[\Sigma_i,\Sigma_{j}]
\eeq
which act on the spinor fields according to
\beq
\begin{split}
\Psi(x) \rightarrow \Psi'(x)&=\exp\left(i\omega^{ij} Y_{ij}\right) \Psi(x) \\
\overline{\Psi}(x) \rightarrow \overline{\Psi'}(x)&=\overline{\Psi}(x) \exp\left(-i\omega^{ij} Y_{ij}\right)
\end{split}
\eeq

\section{RG analysis of O$(N_\Sigma)$ model with $\epsilon$ expansion}\label{sec:RGanalysis}
Let's consider the theory in $D=2+\epsilon$ dimensions. The bare Lagrangian is
\begin{align}
\mathcal{L}_{0}=\overline{\Psi}_0 \left(\slashed{\dou}+\mu \boldsymbol{m_0} \cdot \boldsymbol{\Lambda}\right)\Psi_0
-g_0 \mu^{-\epsilon} \sum_{i}\left(\overline{\Psi}_0\Sigma_i \Psi_0\right)^2
\end{align}
$\boldsymbol{m_0}$ and $g_0$ are dimensionless.
Defining the renormalized quantities according to 
\beq
\begin{split}
\Psi_0&=Z^{\frac{1}{2}}_{\Psi} \Psi\\
\boldsymbol{m_0}&=Z_m \boldsymbol{m}\\
g_0&=Z_g g
\end{split}
\eeq
The renormalized lagrangian is
\begin{align}
\mathcal{L}_R=Z_{\Psi}\overline{\Psi} \slashed{\dou}\Psi+Z_1\overline{\Psi}\;\mu\boldsymbol{m} \cdot \boldsymbol{\Lambda}\;\Psi 
-Z_2 g\mu^{-\epsilon} \sum_{i}\left(\overline{\Psi}\Sigma_a \Psi\right)^2
\end{align}
\beq
\begin{split}
Z_{\Psi}&=1+A\\
Z_1&=Z_m Z_{\Psi}=1+B\\
Z_2&=Z_g Z^2_{\Psi}=1+C
\end{split}
\eeq
\beq
A=\sum_{n=1}^{\infty}A^{(n)} g^n
\eeq
Similarly, $B$ and $C$ are expanded in powers of $g$.

We use dimensional regularization with $\overline{\text{MS}}$ scheme to calculate the $Z$s. To determine the $Z$s up to $m^{\text{th}}$ loop order, the self-energy diagrams and four-point vertex functions up to $m$ loops must be evaluated. In a renormalizable theory, the logarithmic terms with poles like$\frac{1}{\epsilon^k}\ln{\frac{m^2}{4\pi e^{-\gamma}}}$ should cancel and not enter the $Z$s. This fact can be used to obtain a simplification  where the calculation of four-point vertex functions can be avoided by calculating the self-energy to one higher order in loops and demanding the above criterion. At one loop level, the renormalization is given by
\beq
\begin{split}
Z_{\Psi}&=1+O(g^2)\\
Z_1&=1+\frac{N_{\Sigma}}{\pi\epsilon}g+O(g^2)\\
Z_2&=1+\frac{4N_{\Sigma}+N_{\gamma}-6}{\pi\epsilon}g+O(g^2)
\end{split}
\eeq
To calculate the anomalous dimension of the identity mass, we add the fermion bilinear with identity mass to gap out the single-particle spectrum:
\beq
\mathcal{L}_{0}=\overline{\Psi}_0 \left(\slashed{\dou}+\mu M_0 \right)\Psi_0
-g_0 \mu^{-\epsilon} \sum_{i}\left(\overline{\Psi}_0\Sigma_a \Psi_0\right)^2
\eeq
The renormalization at one loop level is then
\beq
\begin{split}
Z_{\Psi}&=1+O(g^2)\\
Z_1&=1-\frac{N_{\Sigma}}{\pi\epsilon}g+O(g^2)\\
Z_2&=1+\frac{N_{\gamma}-2}{\pi\epsilon}g+O(g^2)
\end{split}
\eeq
\section{Repulsive Heisenberg Model : Schwinger boson MFT}\label{sec:schwingerboson}
In the limit, $t \to 0$ of the Hubbard model (\eqnref{eqn:modifiedmodel}) at half filling, each site of the lattice is singly occupied in the ground state manifold with the spins of electrons at each site emerging as the low energy degrees of freedom. For $t \ll \abs{h}$, the effective low energy hamiltonian is the Heisenberg model with repulsive interactions between the emergent spin degrees of freedom. If the fermion hopping amplitude between two sites is $t$, then the Heisenberg coupling between the spin degrees of freedom at those sites is $J=\frac{t^2}{\abs{h}}$. The effective hamiltonian is:
\beq\label{eqn:Heisenbergmodel}
H_{\text{eff}}=\sum_{I a, I' a'} \frac{t_{I a, I' a'}^2}{\abs{h}} \bS_I \cdot \bS_J
\eeq
Let us call the couplings on the red links as $J_1= J=\left(1+r\right)^2\frac{t^2}{\abs{h}}$, and those on the black links as $J_2=\lambda J=\left(1-r\right)^2\frac{t^2}{\abs{h}}$ where $\lambda=\frac{\left(1-r\right)^2}{\left(1+r\right)^2}$. In the limit $r \to -1$ or $r \to 1$, the system is in the plaquette valence bond phase, while at $r=0$, we expect it to be in the N\'eel AF phase. We would like to know if AF order survives after introducing $r \neq 0$. We investigate this using the Schwinger boson MFT technique.

The Schwinger boson representation of spin $S$ at site $Ia$ is given by
\beq\label{eqn:schwingerboson}
\bS_{Ia}=\frac{1}{2} b_{Ia\mu}^{\dag} \sigma_{\mu \nu} b_{Ia \nu}
\eeq
$\mu$, $\nu= \, \up$ or $\down$ (two flavors of bosons per site), and the boson Hilbert space is constrained by
\[\sum_\mu b_{Ia\mu}^{\dag} b_{Ia\mu}=2S\]
After writing the Heisenberg hamiltonian in terms of Schwinger bosons, to do mean-field decomposition in the antiferromagnetic channel, we need the identity\cite{YBKim}
\beq
\sum_{k=x,y,z} \sigma_{\mu \nu}^{k} \sigma_{\rho \sigma}^{k}=-2 \epsilon_{\mu\rho}\epsilon_{\nu\sigma}+\delta_{\mu\nu}\delta_{\rho\sigma}
\eeq
With this, the hamiltonian can be written in a form suitable for mean-field decomposition:
\beq
\bS_{Ia} \cdot \bS_{I' a'}=-\frac{1}{2} A_{I a, I' a'}^{\dag} A_{I a, I' a'}+S^2
\eeq
where
\beq
A_{I a, I' a'}=b_{Ia\mu} \epsilon_{\mu\nu}b_{I' a'\nu}
\eeq
We would have to introduce a Lagrange multiplier at each site, $\mu_{Ia}$ into the hamiltonian as $\sum_{Ia} \mu_I \left(\sum_\alpha b_{Ia\alpha}^{\dag} b_{Ia\alpha}-2S\right)$. In our approximation, we impose the constraint only on the average \cite{Sarker}. Assuming the mean-field values $\expect{A_{I a, I' a'}}=A_1(A_2)$ on the bonds of type $1(2)$, the mean-field hamiltonian per unit cell is
\beq\label{eqn:hbergmft}
\frac{H_{\text{SB,MF}}}{N}=E'(A_1,A_2,\mu)+\frac{1}{N}\sum_{\bk \mu} \bpm b_{\bk\mu}^{\dag}&b_{\bk\bar{\mu}}\epm \mathcal{H}_k^{\text{SB}} \bpm b_{\bk\mu}\\b_{-\bk\bar{\mu}}^{\dag} \epm
\eeq
where 
\beq
E'(A_1,A_2,\mu)=2J\left(A_1^2+\lambda A_2^2\right)-4\mu\left(2S+1\right)
\eeq
\beq
\mathcal{H}_k^{\text{SB}}=\bpm \frac{\mu}{2} & h_{\bk}\\ h_\bk & \frac{\mu}{2} \epm
\eeq
$h_\bk$ is the hopping matrix corresponding to the given bond strengths. The hamiltonian is to be diagonalized by a Bogoliubov transformation of bosons. The operator column object satisfies the algebra
\beq
[b_i,b_j^\dagger]=\eta_{ij}
\eeq
$\eta_{ij}=\sz \otimes \id{4}$. Any transformation we make must preserve this algebra. If $b \to \chi= \mathcal{M} b$ is a transformation, $\mathcal{M}$ should satisfy $\mathcal{M}\eta\mathcal{M}^{\dag}=\eta$. We want $\mathcal{M}^{\dag}\mathcal{H} \mathcal{M}=\Lambda$ to be a diagonal matrix. Thus, $\mathcal{M}^{-1} \eta \mathcal{H} \mathcal{M}=\Lambda \eta$. Diagonalizing the matrix $\eta \mathcal{H}_k$ to obtain the eigenvalues of the hamiltonian \eqnref{eqn:hbergmft}, we have:
\beq
\frac{H_{\text{MF}}}{N}=E^0(A_1,A_2,\mu)+\sum_{\bk, s=\pm, p} \omega_{\bk s} \; \; \chi^\dagger_{\bk s, p}\chi_{\bk s ,p}
\eeq
where
$E^0(A_1,A_2,\mu)=E'(A_1,A_2,\mu)+\frac{4}{N}\sum_{\bk, s=\pm} \omega_{\bk s}$ and $\omega_{\bk \pm}=\sqrt{\left(\frac{\mu}{2}\right)^2-\epsilon_{\pm}^2(\bk)}$. $\epsilon_{\pm}^2(\bk)$ are the eigenvalues of the hopping matrix $h_{\bk}$ with $J_1 A_1$ and $J_2 A_2$ as the hopping amlpitudes.
\begin{align}
\nonumber
\epsilon_{\pm}^2(\bk)=&\frac{J^2}{8} \Big(A_1^2 +\lambda^2 A_2^2 + \lambda A_1 A_2 g(\bk) \Big. \\
&\pm f_x(A_1,A_2,\bk) f_y(A_1,A_2,\bk) \Big. \Big)
\end{align}
with the identification 
\begin{align}
\nonumber
f_i(a,b,\bk)&=\sqrt{a^2+\lambda^2 b^2+2\lambda a b \cos{k_i}}\\
g(\bk)&=\cos{k_x}+\cos{k_y}
\end{align}
The particle equation is given by $\frac{\partial E^0}{\partial \mu}=0$ and the ground state energy is to be minimized with respect to the mean-field values: $\frac{\partial E^0}{\partial A_1}=\frac{\partial E^0}{\partial A_2}=0$. Denoting $A_1=A$, the ratio of mean field values $\frac{A_2}{A_1}=v$, and $\td{\mu}=\frac{\mu}{J A}$
\begin{align}
\nonumber
\epsilon_{\pm}^2(\bk)&=\frac{J^2 A^2}{8}\Big(1+\lambda^2 v^2 + \lambda v g(\bk) \Big. \pm f_x(1,\lambda v) f_y(1,\lambda v) \Big. \Big)  \\
&:=  \frac{J^2 A^2}{4} \Omega_{\pm}^2(\bk)
\end{align}
In terms of the dimensionless quantities defined above, the particle equation can be written as
\begin{align}\label{eqn:particleeq}
    \frac{1}{2N} \sum_{\bk s} \frac{\td{\mu}}{\td{\mu}^2-\Omega_{s}^2(\bk)}-\left(2S+1\right)=0
\end{align}
Energy minimization conditions are given by
\begin{equation}
\resizebox{0.9\linewidth}{!}{$A=\frac{1}{8} \sum_{\bk, s=\pm} \frac{2+ \lambda v g(\bk)+s\left(\sum_{i}^{x,y}\sqrt{\frac{f_{\overline{i}}\left(1,\lambda v, \bk\right)}{f_i\left(1,\lambda v,\bk \right))}}\left(1+\lambda v \cos{k_i}\right)\right)}{\td{\mu}^2-\Omega_{s}^2(\bk)}$}
\end{equation}
\begin{equation}
\resizebox{0.9\linewidth}{!}{$v A=\frac{1}{8} \sum_{\bk, s=\pm} \frac{2 \lambda v+ g(\bk)+s\left(\sum_{i}^{x,y}\sqrt{\frac{f_{\overline{i}}\left(1,\lambda v, \bk \right)}{f_i\left(1,\lambda v, \bk \right))}}\left(\lambda v+ \cos{k_i}\right)\right)}{\td{\mu}^2-\Omega_{s}^2(\bk)}$}
\end{equation}
Taking the ratio of the two equations, we get an equation of the form
\begin{align}\label{eqn:vsolve}
    v=v_{\text{solve}}(\td{\mu},v;\lambda)
\end{align}
$\mu \geq2 \left(\epsilon_{\pm}^2(\bk)\right)_{\text{max}}=J A (1+\lambda v) $. If \eqnref{eqn:particleeq} has no solution for $\td{\mu} \geq 1+\lambda v$, then the ground state is a bose condensate of the $\chi$ bosons. We want to identify the range of parameter $r$ for which bose condensate and hence AF order occurs. Assuming that a condensate forms and setting $\td{\mu}=\td{\mu}_c=1+\lambda v$, \eqnref{eqn:vsolve} is solved and the resulting solution is plugged in \eqnref{eqn:particleeq} to check for particle deficiency. If the particle equation gives particle deficiency, it indicates that Bose condensate is formed.

We find that for $S=\frac{1}{2}$, bose condensate occurs and the ground state is AF for $\lambda_c<\lambda<\lambda_c^{-1}$, where $\lambda_c \approx 0.325$. i.e., AF order survives for $-0.27<r<0.27$ and a transition to the VBS occurs upon increasing $\abs{r}$.

\clearpage

\fi 




\end{document}